\documentclass{emulateapj}
\usepackage{amsmath}
\usepackage{graphicx}
\usepackage{amsfonts}
\usepackage{natbib}
\usepackage{color}
\usepackage{epstopdf}
\usepackage{graphicx}

\begin{document}

\title[$\gamma$-ray QPO in PKS 2155-304]{Revisiting Quasi-Periodic modulation in $\gamma$-ray Blazar PKS 2155-304 with \emph{Fermi} Pass 8 data}
\author{Peng-fei~Zhang\altaffilmark{1,3}, Da-hai~Yan\altaffilmark{2,4,5}, Neng-hui~Liao\altaffilmark{1}, Jian-cheng~Wang\altaffilmark{2,4,5}}
\altaffiltext{1}{Key Laboratory of Dark Matter and Space Astronomy, Purple Mountain Observatory, Chinese Academy of Sciences, Nanjing 210008, China; zhangpengfee@pmo.ac.cn}
\altaffiltext{2}{Yunnan Observatory, Chinese Academy of Sciences, Kunming 650011, China; yandahai@ynao.ac.cn}
\altaffiltext{3}{Key Laboratory of Astroparticle Physics of Yunnan Province, Yunnan University, Kunming 650091, China}
\altaffiltext{4}{Center for Astronomical Mega-Science, Chinese Academy of Sciences, 20A Datun Road, Chaoyang District, Beijing, 100012, China}
\altaffiltext{5}{Key Laboratory for the Structure and Evolution of Celestial Objects, Chinese Academy of Sciences, Kunming 650011, China}

\begin{abstract}

We examine the gamma-ray quasi-periodic variability of PKS 2155-304 with the latest publicly available \emph{Fermi}-LAT Pass 8 data
which covers the years from 2008 August to 2016 October.
We produce the light curves in two ways, i.e., the exposure-weighted aperture photometry
and the maximum likelihood optimization.
The light curves are then analyzed by using Lomb-Scargle Periodogram (LSP) and Weighted Wavelet Z-transform (WWZ); and the results reveal a significant quasi-periodicity with a period of $1.74\pm0.13$ years and a significance of $\sim$ 4.9 $\sigma$.
The constraint of multifrequencies quasi-periodic variabilities on blazar emission model is discussed.

\end{abstract}

\bigskip
\bigskip


\section{INTRODUCTION}
\label{sec:intro}

Blazars are a class of radio-loud active galactic nucleis (AGNs) who aim their jets almost
directly at Earth \citep{Urry1995}. Blazar emission extends from MHz radio frequencies
to TeV gamma-ray energies. The multiwavelength radiations from blazars are dominated by the non-thermal radiations from their jets.
Besides the two-bump spectral energy distribution (SED), erratic variability at all wavelengthes is the distinguished feature of blazars emission, which is an important tool for investigating physics of blazars.

Blazar variability is usually aperiodic.
Accompanying the data of blazars increasing with technological advancements of astronomical telescopes,
a few of possible quasi-periodic variabilities of blazars with a broadband of timescales were reported \citep[e.g.,][]{Urry2011,Gupta2014}.
Naturally, these findings are mainly in optical band because of the abundant optical data
\citep[e.g.,][]{Bai1998,Bai1999,Fan2000,Xie2008,Li2009,King2013,Bhatta2016}.
Among of them, the most persuasive case is OJ 287 whose optical quasi-periodic cycle is $\sim$ 12 yeasrs \citep{Kidger1992,Valtonen2006}.
\citet{Zhang2014} reported a quasi-periodic variability of optical emission in PKS 2155-304
\citep[redshift $z$ = 0.116;][]{Falomo1993} with a timescale of 317$\pm$12 days.

Up to now, the Large Area Telescope on the \emph{Fermi Gamma-ray
Space Telescope} has been collecting data for over eight years, which enables the finding of quasi-periodic
variability with timescales of a few years in gamma-ray flux \citep[e.g.,][]{1553,Sandrinelli2016b}.
Interestingly, \citet{Sandrinelli2014} confirmed the optical periodic-variability of PKS 2155-304 by \citet{Zhang2014} with independent data,
and furthermore, they found marginal evidence of its quasi-periodic variability with a period cycle of $\sim$ 619 days in gamma-ray flux with
the \emph{Fermi}-LAT data in the interval 2008 August 6 (Modified Julian Day, MJD, 54684) to 2014 June 9
(56817 MJD). However, \citet{Sandrinelli2014} pointed out that their result is \emph{tentative} due to the limited time interval.

In this paper, we present detailed analysis of the \emph{Fermi}-LAT light curves of PKS 2155-304
in the interval 2008 August 4 (54682.66 MJD) to 2016 October 26.
Our results reveal a $4.9\ \sigma$ significance-level quasi-periodic variability in the gamma-ray flux of PKS 2155-304.
The period cycle is $\sim640$ days, which is consistent with the finding of \citet{Sandrinelli2014}. The paper is organized as follows: we describe the LAT data analysis and show the main results in Section 2;
and in Section 3 we provide a summary and some discussion.


\section{Observations and Data reduction}
\label{sec:Observations}

The Large Area Telescope \citep{Atwood2009,Abdo2009} on board Fermi satellite is
an electron-positron pair production telescope, and it is sensitive to Gamma-ray photons from $\sim$20 MeV to
$\sim$ 500 GeV \citep{Atwood2009}. It has a 2.4 sr field of view, a point spread function $\sim0.8^{\circ}$
and an effective area $\sim$ 8000 $\rm cm^2$ at 1 GeV. This satellite operates in a continuous entire sky survey mode,
every two $\sim$1.6 hours spacecraft orbits.

We use data from the observations from 2008 August 4 to 2016 October 26 (MJD: 54682.66-57687.59)
with a circular region of interest (ROI) of $15^{\circ}$ radius centered on the position of PKS 2155-304.
In the analysis, the new Pass 8 LAT database \citep{Atwood2013} is used with keeping only the SOURCE class
photon-like events (evclass = 128 \& evtype = 3). We exclude events with zenith angle over $90^{\circ}$.
The events from 100 MeV to 500 GeV are selected to improve the point spread function, and to reduce diffuse emissions.
Using Fermi ScienceTools \emph{gtmktime}, we perform a selection to obtain high-quality data in the good time
intervals with the expression (DATA\_QUAL$>$0)\&\&(LAT\_CONFIG==1). We adopt the instrumental response function (IRF) "P8R2\_SOURCE\_V6".
We use files gll\_iem\_v06.fit and iso\_P8R2\_SOURCE\_V6\_v06.txt to model the two background templates including
Galactic and extra-galactic diffuse emissions. To reduce and analysis the database,
we use the Fermi Science Tools version v10r0p5 package and follow the standard procedure as
in the data analysis thread which is provided in Fermi Science Support Center (FSSC)
\footnote{http://fermi.gsfc.nasa.gov/ssc/data/analysis/scitools/}.

The binned maximum likelihood method is adopted to determine the spectral parameters and flux intensity.
The events are extracted from a $20^{\circ}\times20^{\circ}$ square ROI centered on PKS 2155-304's coordinate with energy over 100 MeV.
We bin the data ten equal logarithmically spaced bins per decade in energy dimensionality with
a spatial pixel size of $0.1^{\circ}\times0.1^{\circ}$. For sources modeling, we fit all the photons with a model file.
The model file is composed of all known 3FGL sources \citep{Acero2015} derived from 4 years of survey data in the ROI;
the two background diffuse emissions are included. The model file are generated with script from user contributions
\footnote{http://fermi.gsfc.nasa.gov/ssc/data/analysis/user/make3FGLxml.py}.
In the model file, the spectral shapes of our target and other gamma-ray sources in the ROI are modeled  by using the same models in 3FGL.
The flux normalization and the spectral parameters of all sources within $5^{\circ}$ are set free;
for the sources within $5^{\circ}$ to $10^{\circ}$ only flux normalization is set free;
and the parameters of other sources located in ROI but beyond $10^{\circ}$ are fixed to the values in 3FGL catalogue,
except for setting free the normalization of significantly variable sources
(i.e., Variability\_Index exceeds the threshold of 72.44) in the ROI.
The Test Statistic value (TS), defined as $\rm TS = -2ln(\mathcal{L}_0/\mathcal{L}_1)$,
is used to determine the source detection significance.

\subsection{Production of light curves}
\label{subsec:make lc}

We produce the light curves of PKS 2155-304 through two methods: the maximum likelihood optimization and
the exposure-weighted aperture photometry\citep{Corbet2007,Kerr2011}.

In the method of maximum likelihood optimization, the light curves are obtained by dividing the data into separate
equal time bins on 10-day timescales. We perform the unbinned maximum likelihood fitting technique for each time bin.
In this case, we use the same parameters of sources as the best-fitting model file, but frozen the parameters of spectrum shape,
 as described in \citet{1553}.
Compared with the binned analysis, the events are selected in a circle ROI of $15^{\circ}$ centered
on the position of PKS 2155-304 and the length of timescale is 10 days for each time bin.
The light curve is presented in Fig. \ref{lk} (upper panel).
For testing the impact of difference of time-length on the power, we also obtain the monthly light-curve using the same method
(see the upper panel of Fig. \ref{lk_1m}).

In the method of exposure-weighted aperture photometry, the probabilities of
every photon in the ROI are calculated with the Fermi tools $gtsrcprob$.
Then we sum the probabilities (rather than the number of photons) within $3^\circ$ radius centered on the position of target using 2.5 days time-bin.
And each time-bin counts is weighted by its relative exposure. The light curve is presented in Fig. \ref{ap} (upper panel).

\subsection{Searching for quasi-periodic variability}
\label{subsec:Search qpo}
\subsubsection{Analyses on gamma-ray data}
\begin{figure*}
\centering
	\includegraphics[scale=0.5]{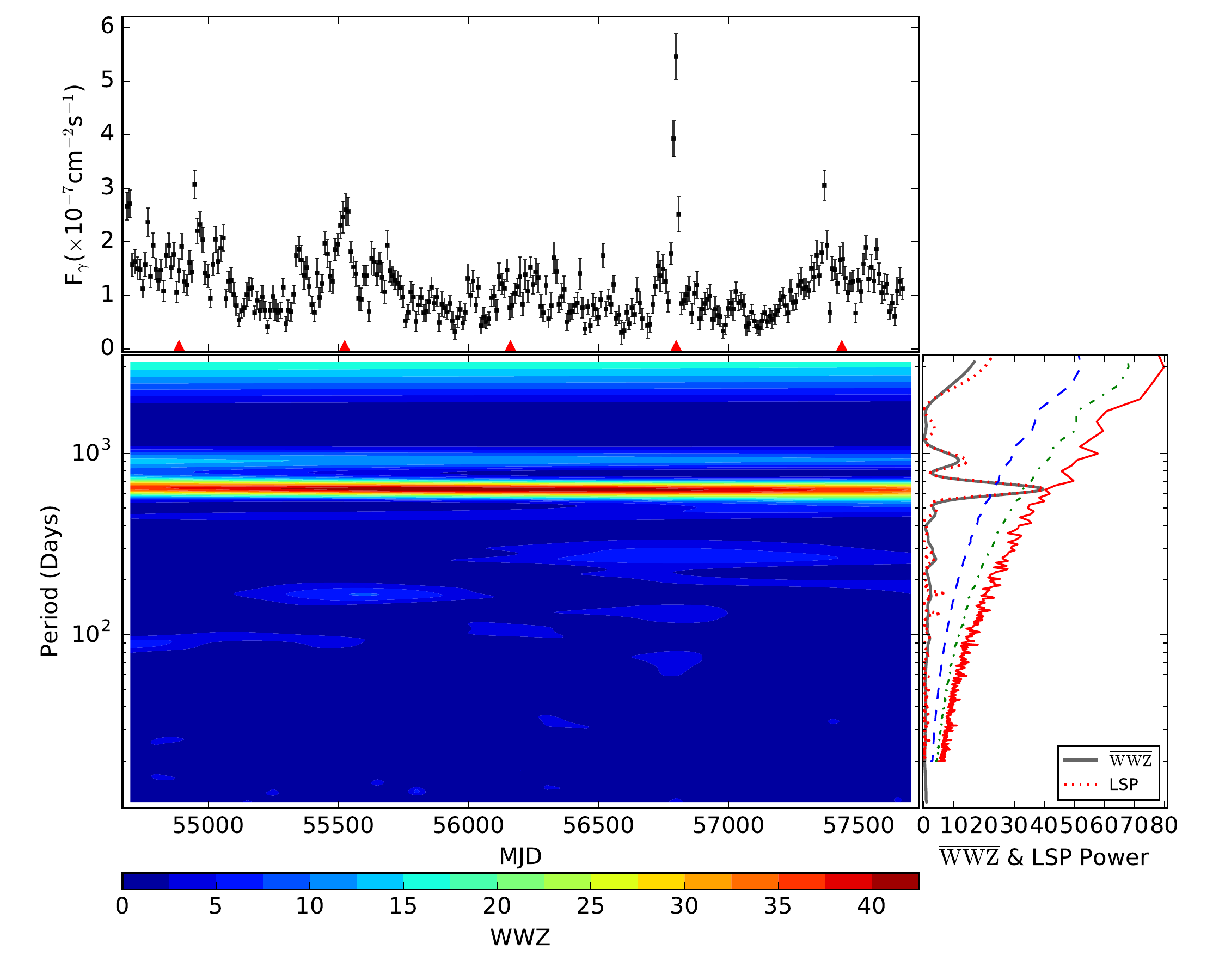}
		\caption{Upper panel: the gamma-ray light curve over 0.1 GeV with a 10 days binning;
		                      the five red uptriangles indicate four period cycles.
		              Left lower panel: the 2D plane contour plot of the WWZ power spectrum of the gamma-ray light curve 
		                           is scaled with color.
		              Right lower panel: the LSP and time-averaged power spectra of the light curve
		                            are shown with red dotted line and black solid line, respectively; the red solid line,
		                            green dash-dotted line and blue dashed line represent the 5 $\sigma$,
		                            4 $\sigma$ and 3 $\sigma$ confidence level, respectively.}
	\label{lk}
\end{figure*}

\begin{figure*}
\centering
	\includegraphics[scale=0.5]{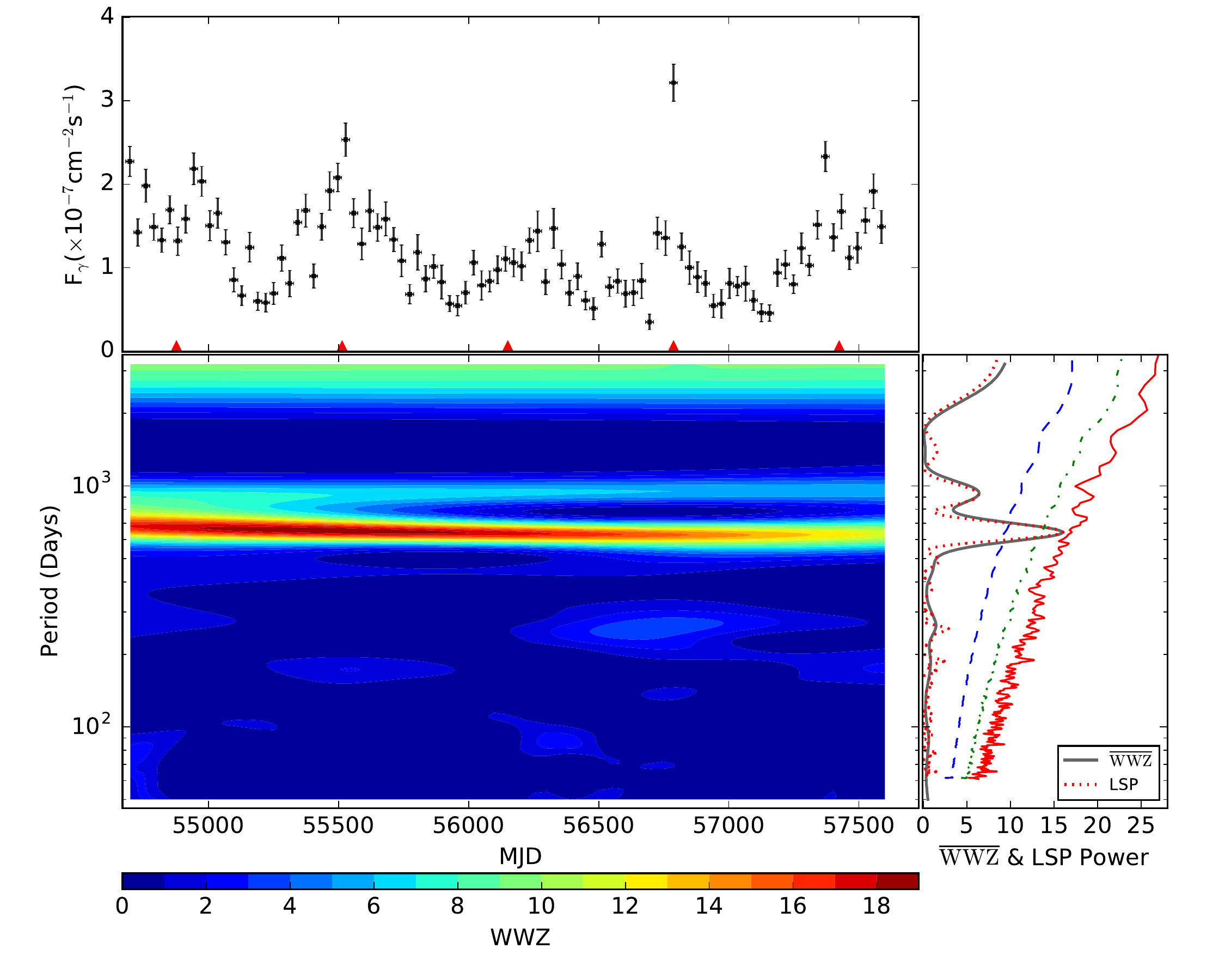}
		\caption{Upper panel: the gamma-ray light curve over 0.1 GeV with a 30 days binning.
		             Left lower panel: the WWZ power spectrum is scaled with color.
		             Right lower panel: the LSP and time-averaged power spectra of the light curve
		                            are shown with red dotted line and black solid line, respectively; the red solid line,
		                            green dash-dotted line and blue dashed line represent the 5 $\sigma$,
		                            4 $\sigma$ and 3 $\sigma$ confidence level, respectively.}
	\label{lk_1m}
\end{figure*}

\begin{figure*}
\centering
	\includegraphics[scale=0.5]{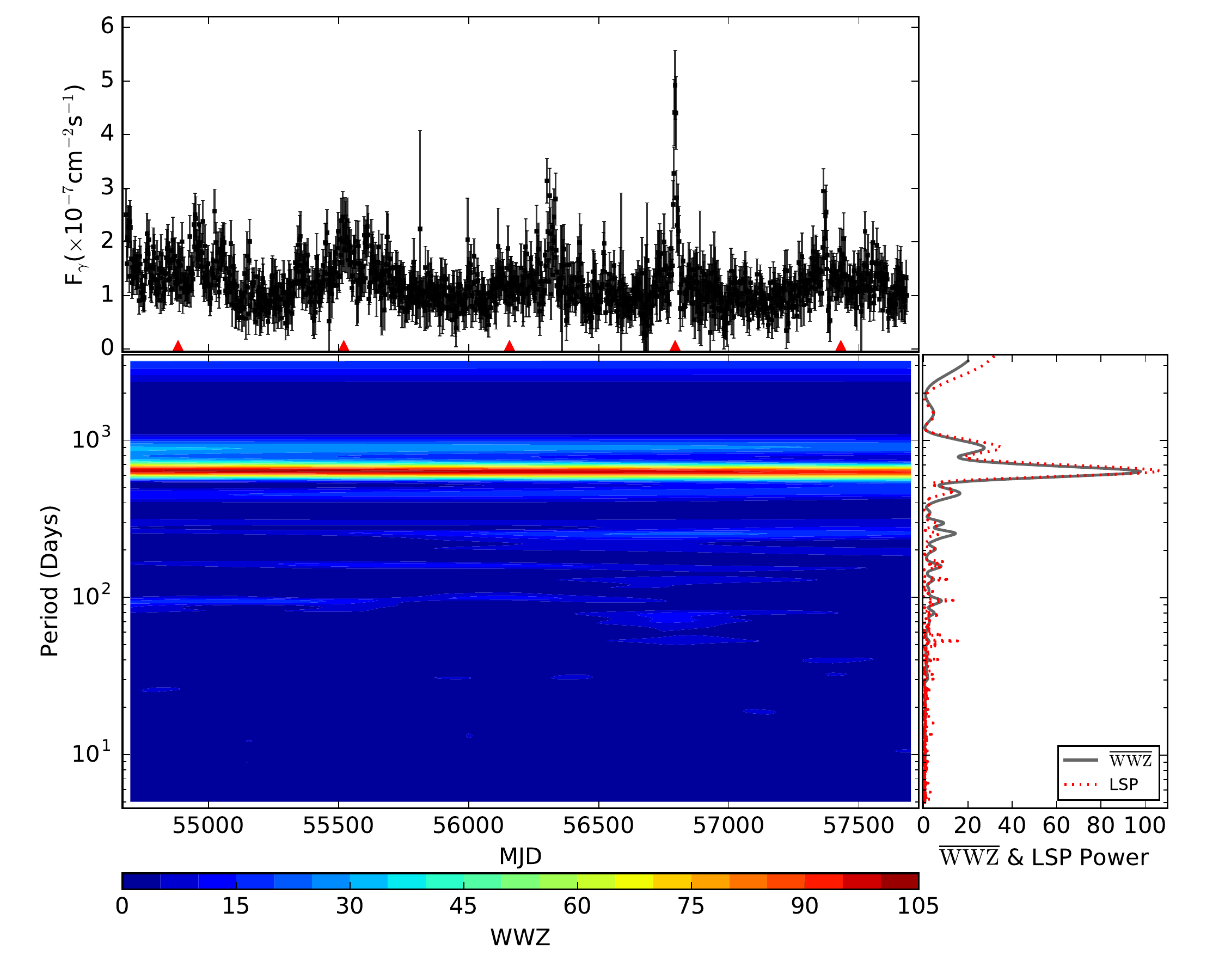}
		\caption{Upper panel: the exposure-weighted light curve over 0.1 GeV with a 2.5 days binning;
		         Left lower panel: the WWZ power spectrum of exposure-weighted light curve is scaled with color. 
		         Right lower panel: the LSP and time-averaged WWZ power spectra of the exposure-weighted light curve are 
		                            shown with red dotted line and black solid line, respectively.}
	\label{ap}
\end{figure*}

We use two methods to search for the quasi-periodic signal, i.e., 
Lomb-Scargle Periodogram \citep{Lomb1976, Scarle1982} and Weighted Wavelet Z-transform \citep{Foster1996}.
The corresponding LSP and time-averaged WWZ powers are shown in the right lower panel in Figs. \ref{lk}-\ref{ap}.
The 2D plane contour plot of the WWZ power spectrum (scalogram) using a Morlet mother function (filled color contour) 
is also shown in Figs. \ref{lk}-\ref{ap} (left lower panel).
All the powers show a clear peak at $\sim640$ days. 
The uncertainty of the period cycle is evaluated with
the half width at half maximum (HWHM) of the Gaussian fitting centered at
the maximum power. Both analyses on the LSP and WWZ powers give a period of $1.74\pm0.13$ year.
We have tested the impact of different time-bins on the position of the peak power, and found that the impact is negligible.

In low-frequency the fluctuations of power are enhanced because of random noise, especially for 
the length of period cycle comparable to the whole light curves. The artificial peak sometimes misidentified as a signal.
The significance of a signal in low-frequency is difficult to be estimated.
For obtaining the robust significance, we simulate light curves based on the obtained power spectral density (PSD) and 
the probability density function (PDF) of observed variation.
The final stimulated light curves therefore have full statistical and variability properties of the gamma-ray flux.
In order to determine the best-fit PSD, we model the power with a power-law function: $P(f)~=~A~f^{-\alpha}~+~c$,
where $\alpha$ is the power-law index
and $c$ marks the Poisson noise level.
The details of the method are provided in \citet{Emmanoulopoulos2013} and \citet{Bhatta2016}.
We simulate $10^{6}$ light curves with the method and evaluate the significance of the signal.
The red solid line, green dash-dotted line and blue dashed line in Figs~\ref{lk}-\ref{ap} represent the 5 $\sigma$, 4 $\sigma$ and 3 $\sigma$ confidence level, respectively.

\begin{figure*}
\centering
	\includegraphics[scale=0.8]{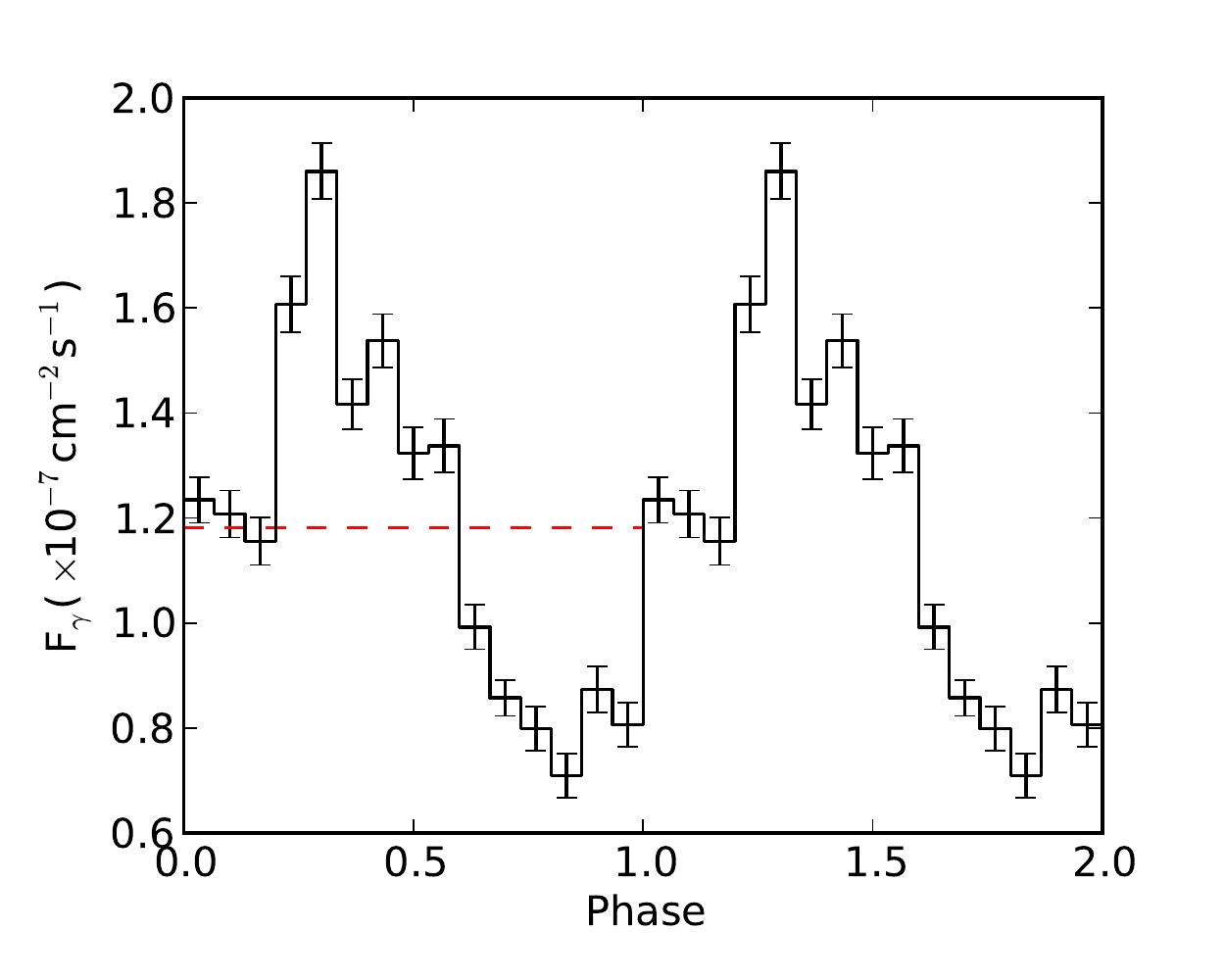}
		\caption{The epoch-folded pulse shape over 100 MeV with 636.7 days period.
		         The red dashed vertical line is the mean flux of the pulse For clarity, we show two period cycles.}
	\label{flod_lc}
\end{figure*}

We fold the gamma-ray photons by using phase-resolved likelihood analysis method with the 636.7 days period.
The folded light curves with phase zero corresponding to MJD 54682.66 is shown in Fig. \ref{flod_lc}. 
We fit the phase-resolved light curves with a constant-flux,
and derive the reduced $\rm \chi^2_{\rm red}=767.5/14$. Therefore we can claim a significant variability of
the phase-resolved light curves. Furthermore this result confirms the signal again.
Clearly, the amplitude of gamma-ray flux varies with phase.

\subsubsection{Analyses on X-ray data}
\label{subsec:xRay}
\begin{figure*}
\centering
	\includegraphics[scale=0.6]{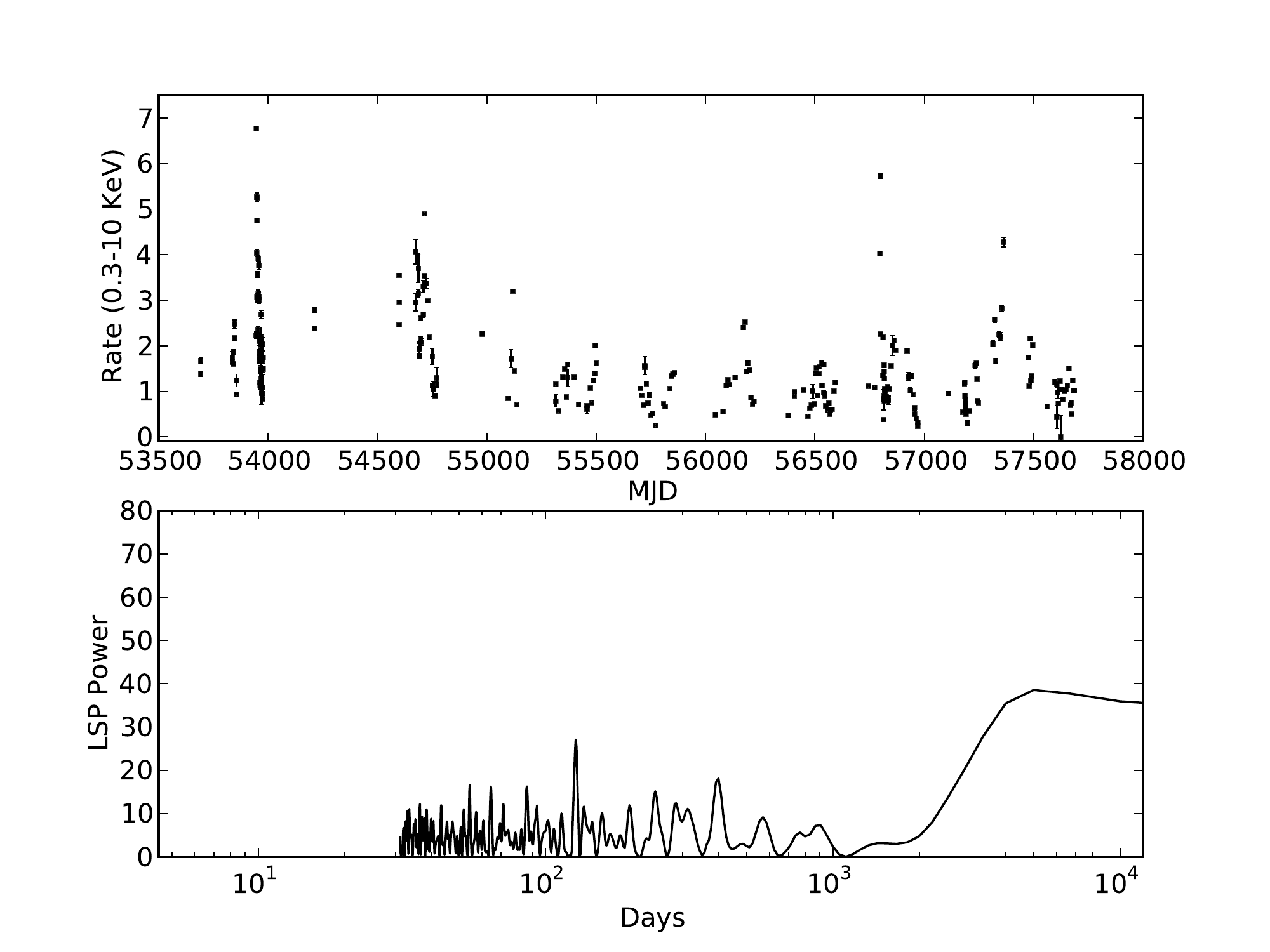}
		\caption{Upper panel: the X-ray (0.3-10 keV) light curve from Swift-XRT of PKS 2155-304.
		              Lower panel: the corresponding LSP power spectrum.}
	\label{lc_x}
\end{figure*}

As far as we know, there is no study on the long term X-ray quasi-period variability of PKS 2155-304 \citep[see][for the study on the short term X-ray quasi-period variability]{Lac}. Therefore,
using same techniques, we analyze a long-term (2005 November 17 to 2016 October 25) X-ray (0.3-10 keV) light curves of PKS 2155-304 from Swift-XRT  \citep{Stroh2013} \footnote{The data have been updated at http://www.swift.psu.edu/monitoring/source.php?source=PKS2155-304}. The results are shown in Fig. \ref{lc_x}.
No significant signal is found in the LSP power spectrum. 


\section{SUMMARY AND DISCUSSION}
\label{sec:summary}

In this paper, we analysed the \emph{Fermi}-LAT data of PKS 2155-304 from 2008 to 2016.
Our results reveal a quasi-periodic variability in gamma-ray flux with a period cycle of $1.74\pm0.13$ years and a
significance of 4.9 $\sigma$, confirming the finding of \citet{Sandrinelli2014}
with much higher significance.

We also analysed the long term X-ray light curve obtained from the observations of Swift-XRT \citep{Stroh2013}.
However, no signifucabt quasi-periodic variability is found, which may be due to insufficient statistics.

\citet{1553} reported marginal evidence for quasi-periodic modulation in $\gamma$-ray flux of PG 1553+113,
which is correlated with the oscillations observed in radio and optical fluxes.
For PKS 2155-304, \citet{Zhang2014} found its quasi-periodic variability
with a quasi-period $\rm T_{1} \sim$ 317 days in optical flux.
This periodicity was confirmed in \citet{Sandrinelli2014} by their data of VRIJHK photometrys.
Moreover, \citet{Sandrinelli2014} revealed a possible quasi-periodic variability of
$\rm 2\times T_{1}$ in gamma-ray flux of PKS 2155-304 with 6-year \emph{Fermi}-LAT data.
\citet{Sandrinelli2016a} found marginal evidence ($\sim3\sigma$) of quasi-periodic variability
with a period cycle $\rm T_2 \sim 280$ days in $Fermi$ gamma-ray light curve of
PKS 0537-441 during a high state extending from 2008 August 10 to 2011 August 27;
however, they found no signal of quasi-period variability in the entire duration of
the \emph{Fermi}-LAT light curve 2008-2015. \citet{Sandrinelli2016a} also found evidence of
quasi-periodic variability with a period cycle $\frac{1}{2}\times\rm T_2$ in optical flux of PKS 0537-441.
However, the optical and gamma-ray quasi-periodic modulations in PG 1553+113 have same period.

So far, three blazars (PG 1553+113, PKS 2155-304 and PKS 0537-441) have been reported having quasi-periodic
variability in gamma-ray fluxes with the significance of $\geq3\sigma$ \citep{1553,Sandrinelli2014,Sandrinelli2016a}.
Among them, the significance of $\sim1.74$ years period gamma-ray cycle for PKS 2155-304 presented in this paper is the highest. 
The three blazars are high-synchrotron-peaked BL Lac objects (HBLs),
whose synchrotron peak frequency is greater than $10^{15}$ Hz \citep{Abdo2010}.
HBLs usually have hard gamma-ray spectra, i.e., the photon spectral index $\Gamma_{\gamma}<2$ \citep{Acero2015}.
The mass of super massive black hole (SMBH) in HBLs is $\sim10^8\ M_\odot$ \citep[e.g.,][]{Woo}.
HBLs are thought to possess a lower accretion rate \citep[e.g.,][]{Ghisellini}. 
On the side of emission mechanism,
the optical-X-ray emission of HBL is the synchrotron radiation from high-energy electrons in the jet,
and the gamma-ray emission is likely produced by synchrotron-self Compton (SSC) process.

The mechanism causing the quasi-periodic
variabilities in blazars is poorly understood \citep{1553},
which should be related to the relativistic jet itself (e.g., jet precession) or to the process
feeding the jet (pulsational accretion flow instabilities).
\citet{1553} have outlined three kinds of models for explaining the quasi-periodic
variabilities in HBLs: (i) pulsational accretion flow instabilities; such instabilities may induce a 
quasi-periodic injection of plasma into jet, which could then produce quasi-periodic
variabilities in jet emissions; (ii) jet precession/rotation, or helical structure; 
in these scenarios Doppler factor changes periodically; (iii) binary SMBH system with a total mass of $\sim10^8\ M_{\odot}$ and a milliparsec separation. Based on the current observations, one cannot determine the physical scenario for the quasi-periodic variabilities in PKS 2155-304. However, on the other side, with the assumption that  the jet optical-$\gamma$-ray emissions are produced in the same region, the above three models predict same period  for optical flux and gamma-ray flux.
If the optical and gamma-ray quasi-periodic modulations in PKS 2155-304 are real, 
the different optical and gamma-ray periods in PKS 2155-304 could challenge the most popular one-zone emission model.  
We would like to suggest that multifrequencies periodic modulations in blazars may put extra constraint on blazar emission models.

We thank the referee for the very helpful suggestions which have significantly improved our manuscript.
We acknowledge the financial support from the 973 Program of China under grant 2013CB837000,
the National Natural Science Foundation of China (No.11573060), Key Laboratory
of Astroparticle Physics of Yunnan Province (No.2016DG006), the Strategic Priority Research Program,
the Emergence of Cosmological Structure of the Chinese Academy of Sciences, Grant No. XDB09000000.


\bibliography{ApJ}

\clearpage
\end{document}